# Formation of Anti hydrogen ion in Positronium - Antihydrogen collision with screened Coulomb Potential

**Dipali Ghosh[1] and C. Sinha[2]**

[1]Michael Madhusudan Memorial College, Durgapur, West Bengal, India, [2] Indian Association For The Cultivation Of Science, Jadavpur, Kolkata-700032, West Bengal, India.

**Abstract:**

The present theoretical work addresses the study of the formation of a positive Antihydrogen ion in a single charge exchange reaction where a positronium ( ground or excited state) interacts with a ground state anti-hydrogen atom, being embedded in a dense Debye plasma. The calculations are carried out in the frame work of a Coulomb Distorted Eikonal Approximation. The result shows that Ps(3S) is the most interesting state to enhance the $\bar{H}^+$ production at the energy close to the reaction threshold. Regarding the Debye screening, it may be inferred that in lower energy regime, the screening effect enhances the TCS for the 1S and 2S states while reduces the same for the 2P and 3S states of Ps .

**Introduction:**

Of late, studies on different antimatters have been the subject of focus both theoretically and experimentally mainly because the Scientists strongly believe that antimatter might play a major role in the creation and expansion of the Universe. It is a highly debatable open question that why the antimatter is so rare as compared to the matter in the observable universe even though equal amount of antimatter should have been created in the Big Bang, commonly known as the baryon asymmetry problem. Recent technological advances in making powerful accelerators, decelerators, storing and trapping devices at CERN by different Groups [1-15] have now made it feasible to produce antiparticles and antimatters artificially at laboratories although with a limited yield.

The most familiar and useful antimatter is the antihydrogen ($\bar{H}$) atom comprising an antiproton ($\bar{p}$) and a positron ($\bar{e}$), the antimatter counterpart of the hydrogen atom. It is now highly anticipated that $\bar{H}$ holds promise to shed light on the question of the aforesaid baryon asymmetry problem which Antihydrogen, the most stable neutral bound state of antimatter offers itself as the most efficient and ideal probe for the precision measurement studies to test the invariance of the CPT symmetry of the Standard model as well as the gravitational Weak Equivalence principle (WEP) of antimatter. In fact by precise spectroscopic comparisons between hydrogen and anti-hydrogen, one can study various fundamental symmetries between matter and antimatter [16]. Any violation of these symmetries could lead to new physics beyond the Standard Model which demands that H and $\bar{H}$ should have the same spectrum. With a view to achieving this long awaited goal, one needs a significant amount of cold and trapped anti hydrogen atoms in ground state suitable for the high precession spectroscopic studies. In fact, experiments with $\bar{H}$ for the study of matter – antimatter symmetry require ultra cold $\bar{H}$ to reach ultimate precision [17]. Unfortunately the anti-hydrogen atoms are preferentially produced at their highly excited (Rydberg) states at laboratories [11] from which ground states are obtained either through spontaneous decay or by means of different efficient methods of stimulated decay.

With a view to enhancing a significant yield of cold and trapped anti hydrogen atoms in their ground state, different theoretical prescriptions [18-21] were also put forward apart from the experimental attempts [22-25].

Another very potential antimatter has recently drawn attention [26], an anti-hydrogen positive ion ($\bar{H}^+$) (composed of an antiproton and two positrons), the counterpart of negative hydrogen ion. The importance of the $\bar{H}^+$ production channel in the coupled channel (CC) calculations has already been emphasized much before [27]. In a very recent work due to Yamashita etal [18; see also other references cited therein] studied the s wave collisions for the system Ps (1s,2s,3s) + $\bar{H}$, adopting a 4 body CC calculation where they included the $\bar{H}^+$ ion formation as open channel as well as all other competing inelastic channels to study the s wave collision cross sections for the production of antihydrogen positive ion. The authors noted significant enhancement in the formation cross sections for the 3s excited state of PS at energies close to threshold as compared to the other two states (1s,2s) for s wave collision.

Interest in the $\bar{H}^+$ ion production mainly stems from fact that it plays an efficient role of an intermediary for the creation of ultra cold anti hydrogen, essential for the test of matter-antimatter gravity. Apart from this, since the $\bar{H}^+$ can easily be manipulated by electric fields, its use in particle and atomic physics cannot be overemphasized. In fact this ion can be utilized to develop an energy tunable anti hydrogen beam that will be used in atomic collision experiments [11].

In the GBAR (Gravitational Behavior of Antihydrogen at Rest) experiment at CERN [28], a different path was chosen for the production of ultra cold neutral anti hydrogen atom which first combines antiproton with two positrons by 3 body recombination (TBR) to form anti hydrogen ions with positive charge. Since this anti matter ions can easily be manipulated they are brought to micro kelvin temperature using laser cooling technique. The extra positron was then photo detached by a very low energy laser pulse to produce finally an ultra cold anti hydrogen ( micro kelvin) atom. Following this experiment [28], a few of theoretical works [18, 19] were also reported for this particular reaction.

However, in most of the anti hydrogen experiments conducted to date, the experiments were performed in plasma environment using cold and dense positron and antiproton plasmas in mainly Penning traps to produce $\bar{H}$. In fact, plasma techniques have been the cornerstone of experimental work in the area of antimatter physics [29]. Plasma confinement and plasma cooling is very effective to confine antiparticles and plasma cooling is also very effective to obtain ultra cold antimatters. In fact experimentally it has been established that single component plasmas have remarkably good confinement properties. Thus theoretically it is highly desirable and quite worthwhile to study the whole reaction mechanism in a plasma environment so that reasonable comparisons can be made with experiments.

With this aim in mind, the present theoretical work addresses the study of the formation of a positive anti-hydrogen ion in a single charge exchange reaction where a positronium ( ground or excited state) interacts with an anti-hydrogen atom producing a positive anti-hydrogen ion in its ground state and an electron, the whole system being embedded in a dense Debye plasma. The calculations are carried out in the frame work of a Coulomb Distorted Eikonal Approximation (CDEA) [30, 31]. Since the binding energy of the negative hydrogen ion is 0.75 eV which is also the energy of the 3$^{rd}$ excited state of the Ps atom, the anti hydrogen ion formation is highly anticipated to increase with excited states of Ps, particularly the third excited state. With a view to this, the present study concentrates on the ground as well as 3 excited states (2s,2p,3s) of Ps as a first venture.

Now inside a dense plasma, partial shielding by the neighboring charged particles weakens the pure coulomb interaction between two charged particles at large separations, thereby affecting the collision properties, e.g., collision strength, collision cross section etc [32-49]. To our knowledge, the present work is the first quantum mechanical calculations for the production of anti-hydrogen positive ion in a plasma environment.

In the Debye plasma, the Coulomb's law is suppressed by the screening to yield a distance dependent charge Q -> Q e $^{-r/\lambda}$, λ being the screening length, The effective coulomb potential is known in plasma physics as the Debye Huckel potential [50] and is given by $V(r) = -\frac{e^{-\mu r}}{r}$ ( negative sign for attractive case), where µ is called the Debye screening parameter given by $\mu = 1/\lambda$. A smaller value of λ can be associated with stronger screening. This screened coulomb potential effectively reduces the binding energy and pushes the system towards gradual instability with the increase of screening. It will thus be of potential interest to study the effect of this screening on the anti hydrogen ion formation which is invariably supposed to be formed inside plasma environments in the experiments carried out at CERN. With a view to this, the present work also motivates to study the variation of the ion formation cross sections with respect to the variation of the plasma screening parameter. The present model of calculation is suitable for low- intermediate energy range which in fact is relatively much higher (~ 1 – 250 ev) as compared to the extreme low energy( ~mev ) experiments at CERN.

## Theory:

The present study calculates theoretically the formation cross sections of positively charged anti hydrogen ion in positronium - antihydrogen charge exchange collision,

$$Ps(1s, 2s, 2p, 3s) + \bar{H}(1s) \rightarrow \bar{H}^+(1s) + e \qquad (1)$$

in the frame work of post collisional Coulomb modified eikonal approximation (CMEA) governed by the Debye-Huckel potential (DHP) or static screened Coulomb potential (SSCP) of the form (in a.u): $V(r) = \dfrac{e^{-\mu r}}{r}$ \qquad (2) where μ is the screening parameter.

The transition amplitudes for the above process

$$T_{if}(\vec{k}_i, \vec{k}_1) = -\dfrac{\mu_f}{2\pi} \langle \psi_f^-(\vec{r}_1, \vec{r}_2, \vec{r}_3) | V_i | \psi_i(\vec{r}_1, \vec{r}_2, \vec{r}_3) \rangle \qquad \text{(prior form)} \qquad (3a)$$

$$T_{if}(\vec{k}_i, \vec{k}_1) = -\dfrac{\mu_f}{2\pi} \langle \psi_f(\vec{r}_1, \vec{r}_2, \vec{r}_3) | V_f | \psi_i^+(\vec{r}_1, \vec{r}_2, \vec{r}_3) \rangle \qquad \text{(post form)} \qquad (3b)$$

Here $V_i$ and $V_f$ are the initial and final channel perturbations which are the part of the total interaction not diagonalised in the initial and final channel. The parameters $\vec{r}_1$, $\vec{r}_2$ and $\vec{r}_3$ denote the coordinates of the electron, the positron in Ps and the positron of the anti hydrogen atom respectively. $\psi_i^+$ or $\psi_f^-$ in equation (3b) or (3a) is an exact solution of the four body problem satisfying the outgoing or incoming wave boundary condition respectively. In the present work we have adopted the prior form of the transition matrix $T_{if}$ as incorporated in equation (3a). Thus in the present case $\psi_f^-$ (equation 3a) satisfies the equation

$$(H - E)\psi_f^- = 0 \qquad (4)$$

where the total Hamiltonian of the present $Ps(1s, 2s, 2p) + \bar{H}(1s)$ system interacting with the DHP is given by,

$$H = -\frac{\nabla_R^2}{2\mu_i} - \frac{\nabla_{12}^2}{2\mu_{Ps}} - \frac{\nabla_3^2}{2} + \frac{e^{-\mu r_1}}{r_1} - \frac{e^{-\mu r_2}}{r_2} - \frac{e^{-\mu r_3}}{r_3} - \frac{e^{-\mu r_{12}}}{r_{12}} - \frac{e^{-\mu r_{13}}}{r_{13}} + \frac{e^{-\mu r_{23}}}{r_{23}} \quad (5)$$

where $\mu_i$ is the three body reduced mass of the system in the initial channel and $\mu_{Ps}$ is the reduced mass of the Ps atom. The initial channel perturbation $V_i$ in equation (3a) is given by

$$V_i = \frac{e^{-\mu r_1}}{r_1} - \frac{e^{-\mu r_2}}{r_2} - \frac{e^{-\mu r_{13}}}{r_{13}} + \frac{e^{-\mu r_{23}}}{r_{23}} \quad (6)$$

where $\vec{r}_{13} = \vec{r}_1 - \vec{r}_3$, $\vec{r}_{23} = \vec{r}_2 - \vec{r}_3$.

The exact final state wave-function $\psi_f^-$ (occurring in prior form of equation 3a) is approximated in the frame work of Coulomb modified eikonal approximation (CMEA) as

$$\psi_f^- = \Phi_{\bar{H}^+}(\vec{r}_2, \vec{r}_3) \exp\left(-\frac{\pi\alpha_1}{2}\right)\Gamma(1 - i\alpha_1)e^{i\vec{k}_1\cdot\vec{r}_1}{}_1F_1\left[-i\alpha_1, 1, -i(k_1 r_1 + \vec{k}_1\cdot\vec{r}_1)\right] \times$$

$$\exp\left[i\eta_1 \int_z^\infty \left(\frac{1}{r_1} - \frac{1}{|\vec{r}_1 - \vec{r}_2|}\right) dz'\right] \quad (7a)$$

with $\alpha_1 = \frac{1}{k_1}$, and $\eta_1 = \frac{1}{k_1}$. The integration variable $z'$ in equation (7a) is the z component of the vector $\vec{r}_1$. To evaluate the phase integral we chose the polar axis in the direction of $\vec{k}_1$, and the result is

$$\psi_f^- = \Phi_{\bar{H}^+}(\vec{r}_2, \vec{r}_3) \exp\left(-\frac{\pi\alpha_1}{2}\right)\Gamma(1 - i\alpha_1)e^{i\vec{k}_1\cdot\vec{r}_1}{}_1F_1\left[-i\alpha_1, 1, -i(k_1 r_1 + \vec{k}_1\cdot\vec{r}_1)\right]$$

$$\times (r_1 + z_1)^{i\eta_1}(r_{12} + z_{12})^{-i\eta_1} \quad (7b)$$

where $z_1$ and $z_{12}$ are the z components of the respective vectors $\vec{r}_1$ and $\vec{r}_{12}$. $\Phi_{\bar{H}^+}(\vec{r}_2, \vec{r}_3)$ in equation (7a) and (7b) corresponds to the $\bar{H}^+$ ion ground state wave function of Chandrasekhar [51] and is given by: $\Phi_{\bar{H}^+}(\vec{r}_2, \vec{r}_3) = \left(\frac{1}{4\pi}\right) N(e^{-\alpha r_2}e^{-\beta r_3} + e^{-\beta r_2}e^{-\alpha r_3}) \quad (8)$ with N=0.3948, $\alpha = 1.03925$ and $\beta = 0.28309$. The initial channel wave-function $\psi_i$ of equation (3a) is chosen as

$$\psi_i = \phi_{PS}(|\vec{r}_1 - \vec{r}_2|)\,(1s, 2s, 2p)\,e^{i\vec{k}_i \cdot \vec{R}}\phi_{\overline{H}}(\vec{r}_3) \tag{9}$$

where $\vec{R} = \frac{1}{2}(\vec{r}_1 + \vec{r}_2)$

Thus in the frame work of Coulomb distorted eikonal approximation [30] and the Debye-Huckel potential (DHP) or static screened Coulomb potential [50] the expression for $T_{if}^{Prior}$ in equation (3a) takes the form;

$$T_{if}^{Prior} = -\frac{\mu_f}{2\pi}\iiint \Phi_{\overline{H}^+}^*(\vec{r}_2, \vec{r}_3)\exp\left(-\frac{\pi\alpha_1}{2}\right)\Gamma(1+i\alpha_1)e^{-i\vec{k}_1 \cdot \vec{r}_1}{}_1F_1[i\alpha_1, 1, i(k_1 r_1 + \vec{k}_1 \cdot \vec{r}_1)]$$

$$\times (r_1 + z_1)^{-i\eta_1}(r_{12} + z_{12})^{+i\eta_1}\left(\frac{e^{-\mu r_1}}{r_1} - \frac{e^{-\mu r_2}}{r_2} - \frac{e^{-\mu r_{13}}}{r_{13}} + \frac{e^{-\mu r_{23}}}{r_{23}}\right)\phi_{PS}(|\vec{r}_{12}|)\,e^{i\vec{k}_i \cdot \vec{R}}\phi_{\overline{H}}(\vec{r}_3)$$

$$dr_1 dr_2 dr_3 \tag{10}$$

After performing some calculations [30, 48, 52], the amplitude $T_{if}$ in equation (10) is finally reduced to a two dimensional numerical integral. Having obtained $T_{if}$, the single differential cross sections (SDCS), is given by

$$\frac{d^3\sigma}{dE_3 d\Omega_1} = \frac{k_1}{k_i}|T_{if}|^2 \tag{11}$$

The Total cross sections $\sigma$ (TCS) can be calculated by using the standard relations and is obtained by integrating equation (11) over the solid angle of the ejected electron, i.e.,

$$\sigma = 2\pi \int_0^\pi \frac{d\sigma}{d\Omega_1}\sin\theta\, d\theta \quad . \tag{12}$$

In this context it may be mentioned that due to the principle of detailed balance, the transition amplitudes obtained from the post and prior forms should, in principle, be the same if the exact scattering wave functions in the initial or final channel are used, which for a four body problem is a formidable task. In the case of approximate wave functions, the aforesaid post and prior forms might not lead to identical results giving rise to some post–prior discrepancy.

**Results and Discussions:**

In the present work, the single differential cross sections (SDCS) and the total differential cross sections ( TCS) of positively charged anti hydrogen ion formation in positronium - antihydrogen collision under DHP are computed. For this charge transfer process, ground (1S) and three excited states (2S, 2P, 3S) of the incident positronium (Ps) have been considered. Since the present study is made in coplanar geometry, that is $\vec{k}_i$ and $\vec{k}_2$ are in the same plane, the azimuthal angles $\emptyset_i$ and $\emptyset_2$ can assume values $0^0$ and $180^0$.

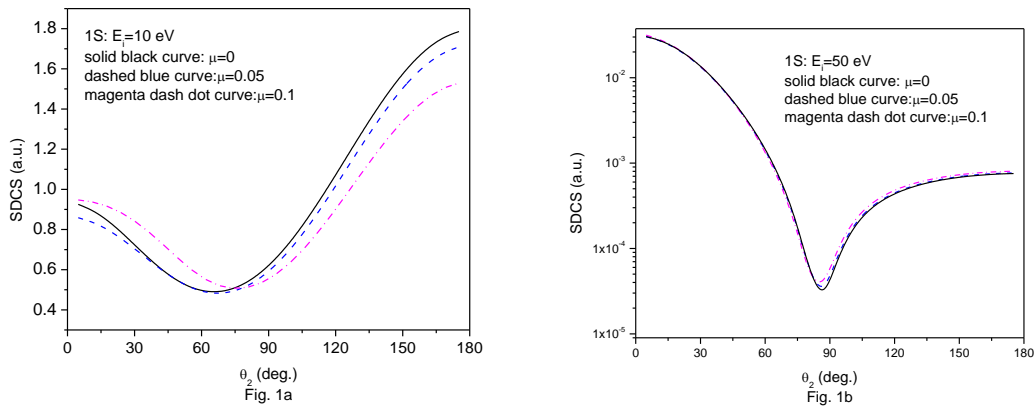

Figure 1: SDCS for positively charged Anti hydrogen formation by Ps (1S) as a function of ejected electron angle ($\theta_2$) for different values of the screening parameter $\lambda$ (in $a_0^{-1}$) and for three different incident energies. (a) $E_i = 10$ eV, (b) $E_i = 50$eV. In both the figures (1a- 1b) the black solid curves correspond to unscreened results, blue dashed curves correspond to $\mu = 0.05$ and magenta dash dot curves correspond to $\mu = 0.1$.

Figures 1(a-b) exhibit SDCS (atomic unit) of $\bar{H}^+(1s)$ formation with ejected electron angle ( $\theta_2$ ) in the charge transfer process of PS(1s)- $\bar{H}(1s)$ system. These figures demonstrate the dependence of Debye screening on incident energies and thus two different screening strengths ($\mu = $ 0.05 and 0.1 $a_0^{-1}$) are chosen for two different incident energies. As expected physically, a completely reverse angular behavior is noted in Fig. 1, e.g., a backward angular dominance at low incident energy ( 10 eV) while a forward peaking at higher incident energy (50 eV) . Figure 1 also reveals that the Debye screening is more sensitive to the SDCs for lesser Ps (1s) energies and also depends on the scattering angle at lower energy regime, e.g., the SDCS is higher for higher screening at forward angles while the reverse is true for backward angles. It is also prominent from figures 1a & 1b, that the overall cross sections get reduced with increasing incident energy associated with a notable qualitative change in distributions.

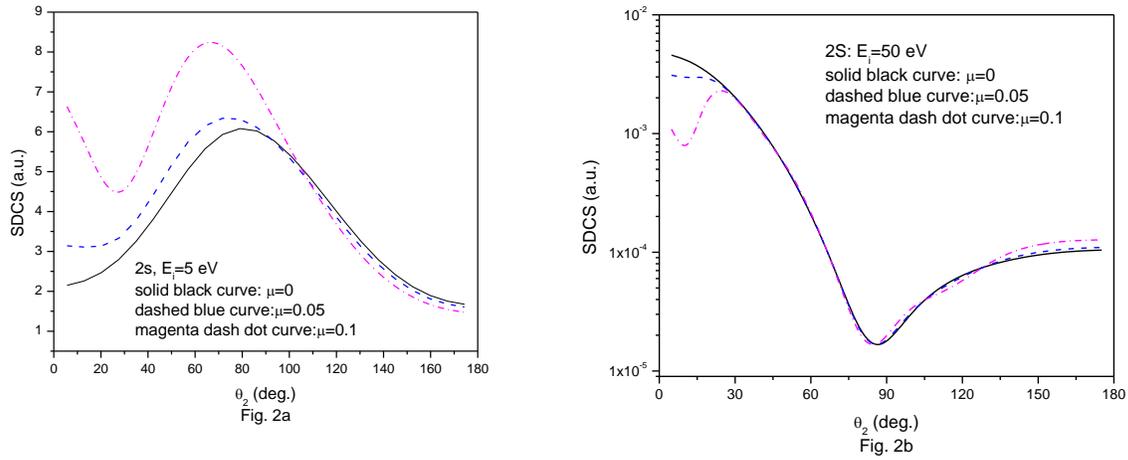

Figure 2. SDCS for Ps (2S) as a function of ejected electron angle ($\theta_2$) for different values of the screening parameter $\mu$ (in $a_0^{-1}$) and for three different incident energies. (a) $E_i = 5$ eV, (b) $E_i = 50$ eV. In both the figures (2a & 2c) the black solid curves correspond to unscreened result, blue dashed curves correspond to $\mu = 0.05$ and magenta dash dot curves correspond to $\mu = 0.1$

Figures 2a & 2b exhibit the SDCS against the ejected electron angle, at different incident energies for the Ps (2S) projectile. From figure 2 it can be seen that magnitude of the SDCS is higher for lower incident energies which is similar to that of Ps (1s) case but the screened SDCS show a different qualitative behavior than that of the unscreened distribution which was absent in figure1(1s). Furthermore, it is noticed that with increasing screening strength, the SDCS surpass the unscreened results in the lower scattering angle region.

Figure 3 exhibits the SDCS (as a function of ejected electron angle) for Ps (2P) projectile to study the effect of screening on the SDCS for different incident energies. The single differential result for Ps (2P) state shows prominent effect of screening at lower incident energy (5 eV). It may be noticed that with increasing incident energy, the screening effect reduces gradually. Comparing figures 1, 2 and 3 it may be inferred that the SDCS (as a function of ejected electron angle) mostly depends on the incident energy and the state of Ps. For the above depicted three states of Ps (1S, 2S, 2P), the SDCS show the maxima at forward ejection angles for higher incident energies while the reverse is true for lower incident energies, e.g., the maxima occur at backward angles.

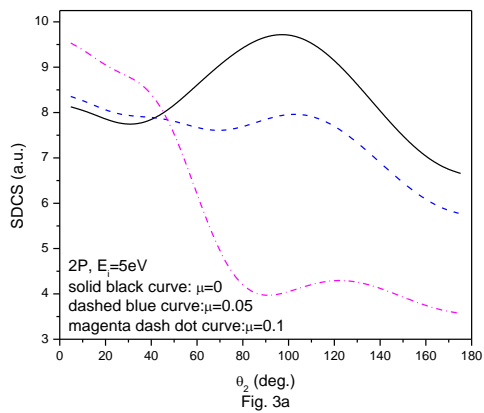
Fig. 3a

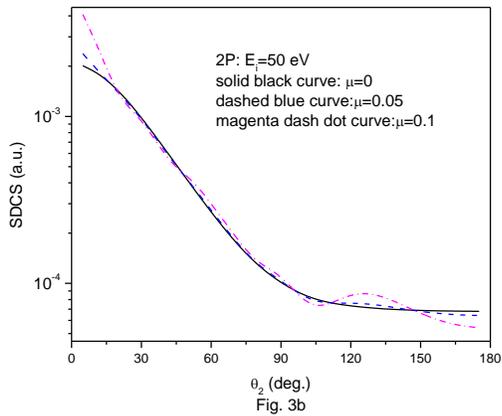
Fig. 3b

Figure 3. SDCS for Ps (2P) as a function of ejected electron angle ($\theta_2$) for two different values of the screening parameter $\mu$ (in $a_0^{-1}$) and for three different incident energies. (a) $E_i = 5$ eV, (b) $E_i = 50$ eV. In both the figures (3a & 3b) the black solid curves correspond to unscreened results, i.e., $\mu = 0$, blue dashed curves correspond to $\mu = 0.05$ and magenta dash dot curves correspond to $\mu = 0.1$.

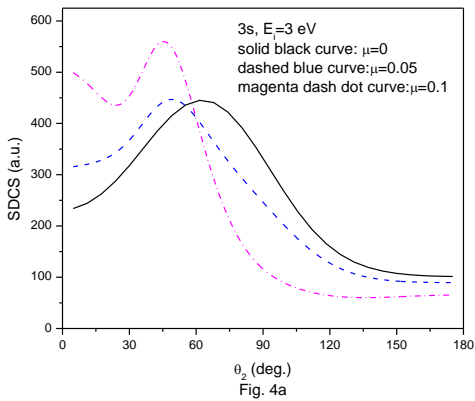
Fig. 4a

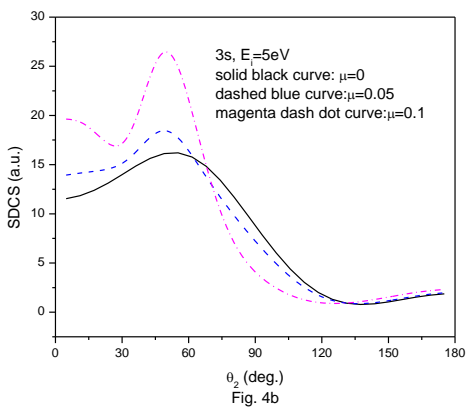
Fig. 4b

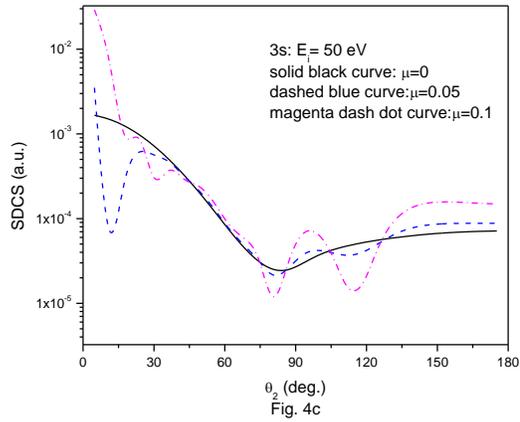

Fig. 4c

Figure 4. SDCS for Ps (3S) as a function of ejected electron angle ($\theta_2$) for different values of the screening parameter $\mu$ (in $a_0^{-1}$) and for three different incident energies. (a) $E_i = 3$ eV, (b) $E_i = 5$ eV, (c) $E_i = 50$ eV. In all the figures (4a- 4c) the black solid curves correspond to screening parameter $\mu = 0$, blue dashed curves correspond to $\mu = 0.05$ and magenta dash dot curves correspond to $\mu = 0.1$.

Similar studies are made for Ps (3S) in figures 4a - 4c. The figures indicate that the screened SDCS go above the unscreened one and it is prominent at forward ejection angles. It is also revealed from the figures that the effect of screening depends on the screening parameters as well as on the incident energies.

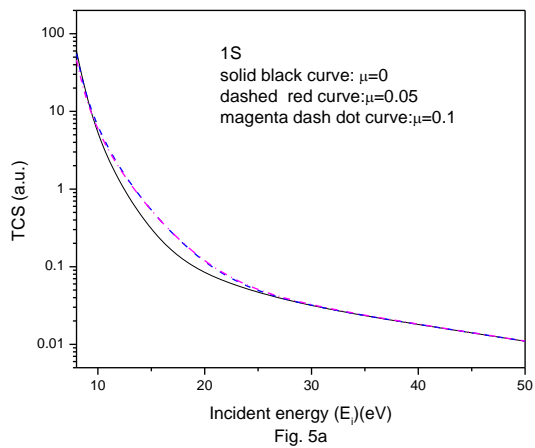

Fig. 5a

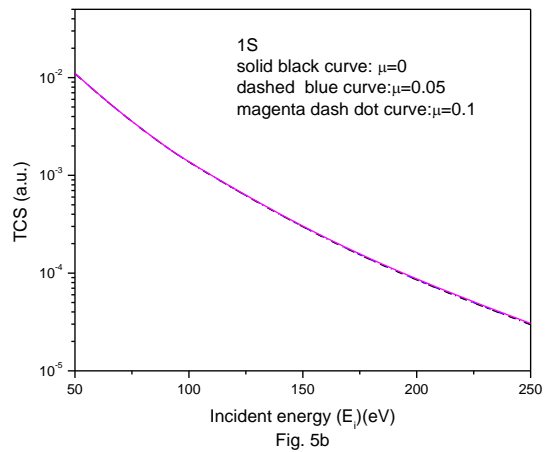

Fig. 5b

Figure 5. TCS for Ps (1S) as a function of incident energy for different values of the screening parameter $\mu$ (in $a_0^{-1}$). (a) $E_i = 7$ eV to 50 eV , and (b) , $E_i = 50$ eV to 250 eV. In the figures (5a & 5b) the black solid curves correspond to screening parameter $\lambda = 0$, blue dashed curves correspond to $\mu = 0.05$ and magenta dash dot curves correspond to $\mu = 0.1$.

Figure 5 exhibits the distribution of TCS against the incident energy of PS (1s) for two different screening parameters. In figure 5a, the incident energy range is from 8 eV to 50 eV , while figure 5b depicts the same for 50 eV to 250 eV. The figure shows that in the energy range ∼ 12 to 23 eV, the screened TCS surpass the unscreened one, but for both very low and high incident energy region, the screening effect is noted to be almost negligible.

**Conclusion**

The present work estimates the formation cross sections of positively charged anti-hydrogen ions from ground state antihydrogen atoms by Ps projectile impact in the presence of a Debye plasma screening which is expected to provide some guidelines to the future experiments. In the absence of any experimental data till now, it is difficult to put this model to the test though there is another theory [20] that was used for the same reaction but for antiproton projectile in vacuum. Despite the discrepancy between theories, it may be noticed that the relative qualitative behavior of the cross sections is more or less the same.